\documentclass[prd,nofootinbib,showpacs,showkeys]{revtex4}

\usepackage{hyperref}
\usepackage{epic,amsmath,graphicx}
\def\mb#1         {\mbox{\boldmath $#1$}}
\def\vmb#1{\vec{\mbox{\boldmath$#1$}}}

\begin{document}

\title{New Heavy-Light Mesons $Q\bar q$}

\author{Takayuki Matsuki}
\email[E-mail: ]{matsuki@tokyo-kasei.ac.jp}
\affiliation{Tokyo Kasei University,
1-18-1 Kaga, Itabashi, Tokyo 173-8602, JAPAN}
\author{Toshiyuki Morii}
\email[E-mail: ]{morii@kobe-u.ac.jp}
\affiliation{Graduate School of Science and Technology,
Kobe University,\\ Nada, Kobe 657-8501, JAPAN}
\author{Kazutaka Sudoh}
\email[E-mail: ]{kazutaka.sudoh@kek.jp}
\affiliation{Institute of Particle and Nuclear Studies, 
High Energy Accelerator Research Organization, \\ 
1-1 Ooho, Tsukuba, Ibaraki 305-0801, JAPAN}

\date{\today} 

\begin{abstract}
We succeed in reproducing the $\ell=1$ $B$ mesons, $B_1(5720)$, $B_2^*(5745)$, and
$B_{s2}^*(5839)$ that are recently reported by D0 and CDF, by using our
semirelativistic quark potential model, which also succeeded in predicting the mass
spectra of the narrow $D_{sJ}$ as well as broad $D_0^*(0^+)$ and $D_1'(1^+)$ particles a
couple of years ago.

Mass of higher excited states ($\ell=1, 2$) of $B$ and $B_s$ mesons, which are not yet
observed, is also predicted at the first order in $p/m_b$ with internal quark momentum
$p$ and the $b$ quark mass $m_b$. We find the corresponding
$B_{sJ}$ are below $BK/B^*K$ threshold and should have narrow decay widths contrary to
most other predictions. Also already established states ($\ell=0$ and $\ell=1$) of
$D$, $D_s$, $B$, and $B_s$ heavy mesons are simultaneously reproduced in good
agreement with experimental data within one percent of accuracy. To calculate these
$D/D_s$ and $B/B_s$ heavy mesons we use different values of strong coupling
$\alpha_s$.
\end{abstract}
\preprint{}
\maketitle

\section{Introduction}
\label{intro}
Recently discovered were narrow meson states, $D_{s0}(2317)$ by BaBar
\cite{BaBar03} and $D_{s1}(2460)$ by CLEO \cite{CLEO03}, both of which were
confirmed by Belle \cite{Belle03}. These are identified as $j^P=0^+$ and $1^+$
of $c\bar s$ excited ($\ell=1$) bound states, respectively, and have caused
a revival of study on heavy meson spectroscopy. Subsequently, another set of
broad heavy mesons, $D_{0}^{*}(2308)$ and $D_{1}'(2427)$ were discovered by
Belle collaboration \cite{Belle04}. Those are identified as $c\bar q$ ($q=u/d$)
excited ($\ell=1$) bound states and have the same quantum numbers $j^P=0^+$ and
$1^+$ as $D_{sJ}$, respectively. 
The decay widths of these excited $D_{sJ}$ mesons are narrow since the masses
are below $DK/D^{*}K$ threshold and hence the dominant decay modes violate the
isospin invariance, whereas those excited $D$ mesons are broad because of no
such restriction as in $D_{sJ}$ cases. More recent experiments reported
by CDF and D0 \cite{CDF_D006} found narrow $B$ and $B_s$ states of $\ell=1$,
$B_1(5720)$, $B_2^*(5745)$, and $B_{s2}^*(5839)$. These are narrow because these
decay through the D-waves.

There are some discussions \cite{Barnes03} that a quark potential model is not
appropriate to describe these new states. In fact, mass spectra and decay widths
of these states could not be reproduced by relativised/relativistic quark potential
models \cite{Godfrey85, Godfrey91, Ebert98}.
In \cite{Godfrey85, Godfrey91}, for instance,
kinetic terms of quarks are taken as positive definite $\sqrt{p^2+m^2}$ and
spin-independent linear-rising confining and short-range Coulomb potentials are
taken into account together with spin-dependent interaction terms symmetric in $Q$
and $\bar q$.
In \cite{Ebert98}, even though degeneracy among members of a spin doublet in the
limit of heavy quark symmetry is taken into account, they obtained masses more than
a hundred MeV larger than the observed $D_{sJ}$ mesons. The nonrelativistic quark
potential model applied both to mesons and baryons has been so successful in
reproducing low lying hadrons (see the review \cite{Mukherjee93}) that people were
puzzled why it did not work for $D_{sJ}$ mesons (atom like mesons).

To understand these heavy meson states, a number of interesting ideas have been
proposed so far. One is an effective Lagrangian approach.
The authors of \cite{Bardeen03, Bardeen94} have proposed the modified
Goldberger-Treiman relation in an effective Lagrangian,
by which the mass difference between two spin doublets
$(0^-,1^-)$ and $(0^+,1^+)$ are reproduced.
This idea is theoretically successful in explaining $D_s$ states but in fail for
$D$ states though experimentally the relation holds.
Other approaches proposed to understand these new states are the QCD sum rule
\cite{Hayashigaki04}, implications of a two-meson molecule
\cite{Barnes03, Szczepaniak03}, four quark system
\cite{Browder04, Terasaki05, Dmitrasinovic05}, lattice calculation
\cite{Bali03} and so on.
The Bethe-Salpeter (BS) equation can be an alternative way
to take relativistic corrections into account.\cite{Habe87, Gara89, Pierro01}
A coupled channel method \cite{Beveren06} is another candidate to
interpret the new heavy mesons though underlying physics is unclear.
These approaches are not yet completely established even though they may be
interesting themselves. Hence, the problem still remains unsolved by these
approaches and is challenging. See the detailed review \cite{Swanson06} on
heavy mesons.

We believe that a quark potential model still powerfully survives if we treat
a bound state equation appropriately, being able to predict not only mass
differences but also absolute values of hadron masses. In fact, long time
ago before the discovery of $D_{s0}(2317)$ by BaBar, two of us (T.M. and T.M.)
\cite{Matsuki97} proposed a new bound state equation for atom-like mesons, i.e.,
heavy mesons composed of a heavy quark $Q$ and a light antiquark $\bar q$, in
which quarks are treated as four-spinor Dirac particles, and Hamiltonian, wave
functions, and eigenvalues are all expanded in $p/m_Q$ so that a light quark is
treated as relativistically as possible and nonrelativistic reduction is made for
a heavy quark. The heavy quark symmetry is taken
into account consistently within a potential model. It is remarkable that the
model could predict the levels of $D_{s0}(2317)$, $D_{s1}(2460)$,
$D_{0}^{*}(2308)$, and $D_{1}'(2427)$ quite successfully even using rather
obscure parameter values due to a small number of data (ground states and only
a few excited states) at the time of publication \cite{Matsuki97}.

Recent experiments on $B$ and $B_s$ (discovery of states of $\ell=1$,
$B_1(5720)$, $B_2^*(5745)$, and $B_{s2}^*(5839)$) by CDF and D0 \cite{CDF_D006}
serves us a good testing ground whether our formulation, semirelativistic quark
potential model, or other formulation also works or not. One of the main purposes
of this paper is to show that our model can also reproduce these $B/B_s$
states within one percent of accuracy compared with the experiments.
Another aim of this paper is to reproduce and predict the whole spectrum of $D/D_s$
and $B/B_s$ heavy mesons, including higher states, $^3D_1$ and $^3D_2/^1D_2$,
which we failed to have reasonal values in \cite{Matsuki97}.

In the previous paper \cite{Matsuki97}, because of the small number of input
data in those days, we searched for {\sl a local minimum} which fits with at least
the lowest ground states and found that the $\chi^2$ analysis led to a
{\em negative} optimal value of $b$ in a scalar potential $S(r)=b+r/a^2$.
However, in this paper we have obtained a {\em positive} $b$ owing to enough
experimental input data this time using the Minuit.
One difference caused by this sign change
appears in light quark mass, $m_q=m_u=m_d$, whose effective mass in the lowest
order Hamiltonian becomes $m_q+b$. This effective mass becomes negative,
$\sim -38$ MeV for $b<0$ in \cite{Matsuki97} and positive, $\sim 86$ MeV in
this paper. Heavy quark masses, $m_c$ and $m_b$, are also affected by this sign
change, i.e., we have obtained a few hundred MeV smaller values in this paper
than those in \cite{Matsuki97}.
Other significant influence of sign change is on the magnitude
of the corrections of lower components of a heavy quark, which are comparable
to upper component contributions in \cite{Matsuki97} ($b<0$), while
in this paper the lower component contributions are comparable to the second
order contributions in $p/m_Q$ as we will see later.
These are the main difference from \cite{Matsuki97} in the way of data analysis
and otherwise we are using the same approach as before.
In addition to a good result for mass levels of $\ell =0$ and $\ell =1$ states of all
$D$, $D_s$, $B$ and $B_s$ mesons, a positive $b$ causes not only obtaining plausible
values for $k=2~(\ell=2)$ states (corresponding to $^3D_1,\; ^3D_2/^1D_2$)
but also obtaining relatively small
degenerate mass $M_0$ in a spin multiplet so that the corrections become
larger in this paper compared with \cite{Matsuki97}.
That is, the sign of $b$ is important to obtain the whole spectra
of heavy mesons even though the lower lying spectra are correctly predicted in
both cases. Another important and different point from the
former paper is that we treat even the light quark masses ($m_u$, $m_d$, and
$m_s$) as free parameters in this paper, while they were given as input in
\cite{Matsuki97}.

In this paper, using the refined parameters obtained from recent
new experimental data, we elaborate computation of the mass spectra of heavy
mesons $D$, $D_s$, $B$, and $B_s$ mesons, including $D_{s0}(2317)$,
$D_{s1}(2460)$, $D_{0}^{*}(2308)$, $D_{1}'(2427)$, $B_1(5720)$, $B_2^*(5745)$,
and $B_{s2}^*(5839)$.  We find that in this new analysis all of the masses of
not only above mentioned new excited ($\ell=1$) states $j^P=0^+$, $1^+$, and
$2^+$ but also those of other excited ($\ell=1$) states $j^P=1^+$ and $2^+$,
and ground
($\ell=0$) states $j^P=0^-$ and $1^-$ of $D$, $D_s$, $B$, and $B_s$ mesons can
be reproduced well within one percent of accuracy. Furthermore, we {\em predict}
$B_{s0}$ and $B_{s1}$ are below $BK/B^*K$ threshold, which means one can find them
as narrow states and is the same results as in \cite{Matsuki05} although we have
about 100 MeV smaller values for $0^+$ and $1^+$ states. This is contrary to most
other quark potential model predictions as in $D_{sJ}$ particles.

The main difference between our treatment of the quark potential model and others
is that we have systematically expanded interaction terms in $p/m_Q$ and that we
have taken into account lower components of a heavy quark in a bound state.
It turns out that {\sl systematically expanded interaction terms are quite different
from those of nonrelativistic and/or relativised models} except for \cite{Pierro01}.
In \cite{Pierro01} they used the BS equation and projected heavy quark sector on
positive states, otherwise their final formulation is very close to ours.
Another large difference is the magnitude of a light quark mass, i.e., other models
including \cite{Pierro01} use a constituent quark mass while we got much smaller
values, e.g., $m_u, m_d \sim 10$ MeV and $m_s \sim 100$ MeV.

The paper is organized as follows. 
In the next section, we briefly review our model, where some necessary formulas
are presented for readers' understanding and a discussion on a convenient
quantum number $k$ is given. In Sec. \ref{result}, numerical results are
presented in Tables \ref{Dmeson} and \ref{Dsmeson} for the mass levels of $D$
and $D_s$ mesons including their excited states together with the most optimal
values of parameters in Table \ref{parameter}, where we find that calculated
masses are in good agreement with all existing experimental data. Those of
$B$ and $B_s$ mesons are also given in the same section, Tables \ref{Bmeson} and
\ref{Bsmeson}. To show smallness of the second order corrections, we
calculate $D$ meson mass spectra with the second order corrections and give them
in Table \ref{Dmeson2nd}. Section \ref{summary} is devoted to conclusion and
discussion, including the reason why our semirelativistic quark potential
model approach gives good results compared with others.

\section{Brief Review of our Semirelativistic Model}
\label{formalism}
Let us start with an effective two-body Hamiltonian for a system of a heavy
quark $Q$ with mass $m_Q$ and a light antiquark $\bar q$ with mass $m_{\bar q}$
\cite{Fermi49, Bethe55},\footnote{Miyazawa and Tanaka \cite{Miyazawa92} pointed
out that a relativistic bound state equation having this effective Hamiltonian
can be derived with a boundary condition in relative time that is different from
that of the Bethe--Salpeter equation.}
\begin{eqnarray}
H&=&(\vec{\alpha}_{\bar q}\cdot \vec{p}_{\bar q} 
+ \beta_{\bar q} m_{\bar q}) 
+ (\vec{\alpha}_Q\cdot \vec{p}_Q 
+ \beta_Q m_Q) 
+ \beta_{\bar q} \beta_Q S(r) \nonumber \\
&&+ \left\{ 1-\frac{1}{2}\left[\vec{\alpha}_{\bar q}\cdot
\vec{\alpha}_Q + 
(\vec{\alpha}_{\bar q}\cdot \vec{n})
(\vec{\alpha}_Q\cdot \vec{n})
\right]\right\}V(r), 
\label{total_H}
\end{eqnarray}
where $S$ and $V$ are a scalar confining potential and a vector one-gluon
exchange Coulomb potential with transverse interaction and $\vec{n}=\vec{r}/r$
is a unit vector along the relative coordinate between $Q$ and $\bar q$. $S$
and $V$ are given explicitly as
\begin{equation}
S(r)=\frac{r}{a^2}+b, ~~~V(r)=-\frac{4}{3}\frac{\alpha_s}{r},
\label{potential}
\end{equation}
where $a$ and $1/b$ are parameters with length dimension and $\alpha_s$ is a
strong coupling constant. Since a heavy quark $Q$ is sufficiently heavier than
a light antiquark $\bar q$ which is orbiting $Q$, it is reasonable to apply the
Foldy--Wouthuysen--Tani (FWT) transformation \cite{Foldy50} to the heavy quark
related operators in $H$ of Eq.(\ref{total_H}). One should notice that the FWT
is introduced so that free kinetic terms of a heavy quark become $\beta_Q m_Q$
and one can expand the interaction terms in terms of $p/m_Q$ systematically,
\footnote{There is another way of expanding the Hamiltonian. See \cite{MS97}
where the Bloch method \cite{Bloch} is used to expand the effective Hamiltonian
in $p/m_Q$, which is reproduced in Appendix B.} which is equivalent to a
nonrelativistic reduction of the
original model. By doing this way, we can get an eigenvalue equation for an atom-
like bound state $Q\bar q$ \cite{Matsuki97} \footnote{There is a pioneering
work of this approach
\cite{Morishita88}, where
expanded terms of the Hamiltonian in $p/m_Q$ are treated differently from ours.}
\begin{equation}
(H_{\rm FWT}-m_{Q}) \otimes\psi_{\rm FWT} =\tilde{E}\psi_{\rm FWT},
\label{eigen_FWT}
\end{equation}
with
\begin{equation}
H_{\rm FWT}-m_{Q}=H_{-1}+H_{0}+H_{1}+H_{2}+\cdots , 
\label{H_mQ}
\end{equation}
where $\tilde{E}=E-m_Q$ ($E$ being the bound state mass of $Q\bar q$) is the
binding energy and a notation $\otimes$ denotes that gamma matrices of a light
antiquark is multiplied from left with the wave function while those of a heavy
quark from right.  $H_i$ in Eq.(\ref{H_mQ}) denotes the $i$-th order expanded
Hamiltonian in $p/m_Q$ and its explicit expressions are presented in
\cite{Matsuki97} and also in the Appendix A. Then, Eq.(\ref{eigen_FWT}) can be
numerically solved in order by order in $p/m_Q$ according to the standard
perturbation method by expanding the eigenvalue and the wave function as
\begin{eqnarray}
  \tilde{E} &=& E-m_{Q}=E_{0}^{a}+E_{1}^{a}+E_{2}^{a}+\cdots , 
  \label{E_mQ} \\
  \psi_{\rm FWT} &=& \psi_{0}^{a}+\psi_{1}^{a}+\psi_{2}^{a}+\cdots , 
  \label{psi_mQ}
\end{eqnarray}
where a superscript $a$ represents a set of quantum numbers, a total angular
momentum $j$, its $z$ component $m$ and a quantum number $k$ of the spinor
operator $K$ defined below in Eq.(\ref{k_quantum}) and a subscript $i$ of
$E_i^{a}$ and $\psi_i^{a}$ stands for the $i$-th order in $p/m_Q$.

In practice, it is reasonable to first solve the $0$-th order non-trivial
equation by variation to get the $0$-th order eigenvalues and their wave
functions and then to estimate the corrections perturbatively in order
by order in $p/m_Q$ by evaluating the matrix elements for those terms.
The 0-th order eigenvalue $E_0^{a}$ and the wave function $\psi_0^{a}$ for the
$Q\bar q$ bound state is obtained by solving the $0$-th order eigenvalue
equation,
\begin{equation}
  H_0 \otimes\psi _0^{a} = E_0^{a}\psi _0^{a}, \qquad
  H_0 =
  \vec{\alpha}_{\bar q}\cdot\vec{p}_{\bar q}+\beta_{\bar q}
  \left(m_{\bar q}+S(r)\right)+V(r).
\label{0th}
\end{equation}
The wave function which has two spinor indexes and is expressed as $4\times 4$
matrix form is explicitly described by
\begin{eqnarray}
  \psi _0^{a} &=& \left(0~~ \Psi _{j\,m}^k(\vec r) \right),
  \label{0thsols1} \\
  \Psi _{j\,m}^k(\vec r) &=& \frac{1}{r}\left( 
\begin{array}{c}
u_k(r)\;y_{j\,m}^k \\ 
iv_k(r)\;y_{j\,m}^{-k} 
\end{array}
\right),
  \label{0thsols2}
\end{eqnarray}
where $y_{j\,m}^k$ being of $2\times 2$ form are the angular part of the wave
functions and $u_k(r)$ and $v_k(r)$ are the radial parts. The total angular
momentum of a heavy meson $\vec J$ is the sum of the total angular momentum of
the light quark degrees of freedom $\vec S_\ell$ and the heavy quark spin
${1\over 2}\vec \Sigma_Q$:
\begin{equation}
  {\vec J} = {\vec S_\ell} +{1\over 2}\vec \Sigma_Q\qquad {\rm with} \quad
  {\vec S_\ell}=\vec L + {1\over 2}\vec \Sigma_{\bar q},
\end{equation}
where ${1\over 2}\vec\Sigma_{\bar q}\ (={1\over 2}
\vec\sigma_{\bar q}\;1_{2\times 2})$ and $\vec L$ are the $4\times 4$ spin and
the orbital angular momentum of a light antiquark, respectively. Furthermore,
$k$ is the quantum number of the following spinor operator $K$ \cite{Matsuki97}
\begin{equation}
  K = -\beta_{\bar q} \left( \vec \Sigma_{\bar q} \cdot \vec L + 1 \right),
  \qquad
  K\, \Psi_{j\,m}^k = k\, \Psi_{j\,m}^k.
  \label{k_quantum}
\end{equation}
Note that in our approach $K$ is introduced for a two-body bound system of
$Q\bar q$, though the same form of the operator $K$ can be defined in the case
of a single Dirac particle in a central potential \cite{JJ}.  One can easily
show \cite{Matsuki04} that the operator $K^2$ is equivalent to
${\vec S_\ell}^2$ and it holds $k = \pm \left( s_\ell + \frac{1}{2} \right)$
or $s_\ell = \left| k \right| - \frac{1}{2}$. $k$ is also related to $j$ as
\begin{equation}
j=|k| ~~{\rm or}~~ |k|-1 ~~~(k\neq 0) . 
\end{equation}

In order to diagonalize the 0-th order Hamiltonian $H_0$ of Eq.(\ref{0th}) in
the $k$ representation, it is convenient to introduce a spinor representation
of $y_{j\,m}^k$ which is given by the unitary transformation
(Eq.(\ref{Unitary}) below) of a set of the spherical harmonics $Y_j^m$ and the
vector spherical harmonics defined by 
\begin{eqnarray}
  \vec Y_{j\,m}^{(\rm L)} = -\vec n\,Y_j^m, \quad
  \vec Y_{j\,m}^{(\rm E)} = {r \over {\sqrt {j(j+1)}}}\vec \nabla Y_j^m,
  \quad
  \vec Y_{j\,m}^{(\rm M)} = -i\vec n\times \vec Y_{j\,m}^{(\rm E)},
\end{eqnarray}
where $\vec n=\vec r/r$. These vector spherical harmonics are nothing but a set
of eigenfunctions for a spin-1 particle. 
$\vec Y_{j\,m}^{(\rm A)}$ (A=L, M, E) are eigenfunctions of ${\vec J}^2$ and
$J_z$, having the eigenvalues $j(j+1)$ and $m$. The parity of these functions
is assigned as $(-1)^{j+1}$, $(-1)^j$, $(-1)^{j+1}$ for A=L, M, and E,
respectively, since $Y_j^m$ has a parity $(-1)^j$. The unitary transformation
is given by 
\begin{equation}
\left( 
\begin{array}{c}
y_{j\,m}^{-(j+1)} \\
y_{j\,m}^j
\end{array}
\right)=U\left( 
\begin{array}{c}
Y_j^m \\ 
\vec \sigma \cdot \vec Y_{j\,m}^{(\rm M)}
\end{array}
\right), \qquad
\left( 
\begin{array}{c}
y_{j\,m}^{j+1} \\
y_{j\,m}^{-j}
\end{array}
\right)=U\left( 
\begin{array}{c}
\vec \sigma \cdot \vec Y_{j\,m}^{(\rm L)} \\
\vec \sigma \cdot \vec Y_{j\,m}^{(\rm E)}
\end{array}
\right), 
\label{Unitary}
\end{equation}
where the matrix $U$ is defined by
\begin{equation}
U={1 \over {\sqrt {2j+1}}}\left( 
\begin{array}{cc}
{\sqrt {j+1}} & {\sqrt j} \\
-{\sqrt j} & {\sqrt {j+1}}
\end{array}
\right). 
\label{eq:app:u}
\end{equation}
Now, substituting Eqs.(\ref{0thsols1}) and (\ref{0thsols2}) into
Eq.(\ref{0th}), one can eliminate the angular part of the wave function,
$y_{j\,m}^k$, from the eigenvalue equation and obtain the radial equation as
follows,
\begin{equation}
\left( 
\begin{array}{cc}
m_q+S+V & -\partial _r+{k \over r} \\
\partial _r+{k \over r} & -m_q-S+V
\end{array}
\right)\left(
\begin{array}{c}
u_k(r) \\
v_k(r)
\end{array}
\right)
= E^k_0\left(
\begin{array}{c}
u_k(r) \\
v_k(r)
\end{array}
\right), 
\label{radial}
\end{equation}
which depends on the quantum number $k$ alone.  This is just the same form as a
one-body Dirac equation in a central potential. Since $K$ commutes with $H_0$
and the states $\Psi _{j\,m}^k$ have the same energy $E_0^k$ for different
values of $j$, these states having a total angular momentum $j=|k|-1$ and $|k|$,
a spin doublet, are degenerate with the same value of $k$.

The parity of the eigenfunction $\psi_0^{a}$ is defined by the upper
(``large'') component of Eq.(\ref{0thsols2}) and hence taking into account the
intrinsic parity of quark and antiquark, the parity of the meson $Q\bar q$ is
given by \cite{Matsuki04}
\begin{equation}
P=\frac{k}{|k|}(-1)^{|k|+1},
\end{equation}
which is equal to the parity $\pi_{\ell}$ of the light degrees of freedom in the
heavy quark effective theory (HQET) as shown in Table I. It is remarkable that
both the total angular momentum $j$ and the parity $P$ of a heavy meson can be
simultaneously determined by $k$ alone. Or the states can be completely
classified in terms of $k$ and $j$ as shown in Table \ref{table}.

\begin{table*}[t]
\caption{States classified by various quantum numbers}
\label{table}
\begin{tabular*}{13cm}{c|@{\extracolsep{\fill}}cccccccc}
\hline
\hline
  \makebox[1.7cm]{$j^P$}   &   $0^-$   &   $1^-$   &   $0^+$   &   
  $1^+$   &   $1^+$   &   $2^+$   &   $1^-$   &   $2^-$   \\
  $k$   &   -1   &   -1   &   1   &   
   1    &   -2   &   -2   &   2   &   2   \\
  $s_\ell^{\pi_\ell}$ & ${1\over2}^-$ & ${1\over2}^-$ & ${1\over2}^+$ & 
  ${1\over2}^+$ & ${3\over2}^+$ & ${3\over2}^+$ & ${3\over2}^-$ & ${3\over2}^-$
  \\
  $^{2s+1}l_j$ & $^1S_0$ & $^3S_1$  & $^3P_0$  & $^3P_1$, $^1P_1$ &
   $^1P_1$, $^3P_1$  & $^3P_2$  & $^3D_1$  & $^3D_2$, $^1D_2$
  \\ \hline 
  $\Psi_j^k$ & $\Psi_0^{-1}$ & $\Psi_1^{-1}$ & $\Psi_0^1$ & $\Psi_1^1$ &
  $\Psi_1^{-2}$ & $\Psi_2^{-2}$ & $\Psi_1^2$ & $\Psi_2^2$ \\
\hline
\hline
\end{tabular*}
\end{table*} 

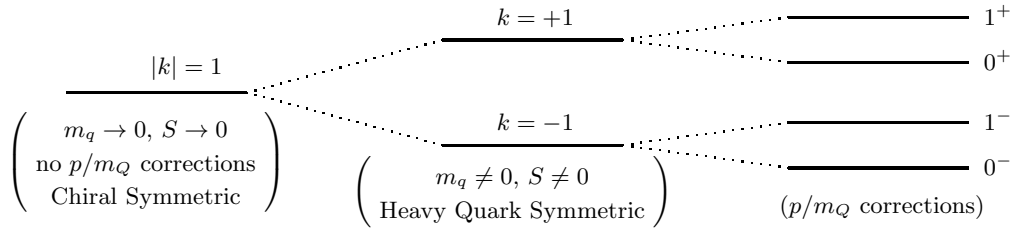
\begin{figure*}[t]
\begin{center}
\begin{picture}(370,90)
\setlength{\unitlength}{0.4mm}
 \thicklines
  \put(5,40){\line(1,0){60}}
  \put(130,57.5){\line(1,0){60}} \put(130,22.5){\line(1,0){60}}
  \put(245,65){\line(1,0){60}} \put(245,50){\line(1,0){60}}
  \put(245,30){\line(1,0){60}} \put(245,15){\line(1,0){60}}
 \thinlines
  \dottedline{3}(65,40)(130,57.5) \dottedline{3}(65,40)(130,22.5)
  \dottedline{3}(190,57.5)(245,65) \dottedline{3}(190,57.5)(245,50)
  \dottedline{3}(190,22.5)(245,30) \dottedline{3}(190,22.5)(245,15)
  \put(33,46){$|k|=1$}
  \put(-15,15){$\left(\begin{array}{c}
              m_q\to0,\, S\to0 \\  {\rm no}~ p/m_Q~ {\rm corrections} \\
              {\rm Chiral~Symmetric}
             \end{array}\right)$}
  \put(148,62.5){$k=+1$} \put(148,27.5){$k=-1$}
  \put(310,62.5){$1^+$} \put(310,47.5){$0^+$}
  \put(310,27.5){$1^-$} \put(310,12.5){$0^-$}
  \put(100,5){$\left(\begin{array}{c}
  			m_q\neq 0,\, S\neq0 \\
  			{\rm Heavy~Quark~Symmetric}
             \end{array}\right)$}
  \put(242,0){($p/m_Q$ corrections)}
\end{picture}
\caption{Procedure how the degeneracy is resolved in our model.}
\label{mass_level}
\end{center}
\end{figure*}

Classification by the quantum number $k$ is convenient to discuss the
relation between the symmetry of the system and the structure of mass levels.
Now, let us look at how masses of $Q\bar q$ mesons are generated in our model.
First, when the light quark mass $m_{\bar q}$ and the scalar potential $S(r)$
are neglected, the $0$-th order Hamiltonian $H_0$ has both a chiral symmetry
and a heavy quark symmetry and the eigenvalues with the same value of $|k|$ are
degenerate in this limit, where all masses of $0^-$, $1^-$, $0^+$, and $1^+$
states with $|k|=1$ are degenerate, as
$m(0^-)=m(1^-)=m(0^+)=m(1^+)=m_Q$.\cite{Matsuki05} Then, by turning on the light quark mass and the scalar
potential, $m_{\bar q}\neq 0$ and $S(r)\neq 0$, the chiral symmetry is broken
and the degeneracy between two spin doublets $(0^-, 1^-)$ and $(0^+, 1^+)$ is
resolved. At this stage, members in spin multiplets are still degenerate due to
the same quantum number $k$. Finally, by adding $p/m_Q$ corrections, the heavy
quark symmetry is broken and the hyperfine splitting occurs as shown in
Fig. \ref{mass_level}.

\section{Numerical Calculation}
\label{result}
To obtain masses of $Q\bar q$ states, we first solve the eigenvalue
equation Eq.(\ref{radial}) by variation to get the lowest eigenvalues and the
wave functions. By taking account of the asymptotic behavior at both
$r\rightarrow 0$ and $r\rightarrow \infty$, the trial functions for $u_k(r)$
and $v_k(r)$ are given by
\begin{equation}
u_{k}(r), ~v_{k}(r)\sim w_{k}(r)\left(\frac{r}{a}\right)^{\lambda}
\exp \left[-(m_{q}+b)r-\frac{1}{2}\left(\frac{r}{a}\right)^{2} \right] , 
\end{equation}
where
\begin{equation}
\lambda=\sqrt{k^{2} - \left(\frac{4\alpha_s}{3}\right)^{2}} , 
\end{equation}
and $w_{k}(r)$ is a finite series of a polynomial in $r$ 
\begin{equation}
w_{k}(r)=\sum_{i=0}^{N-1}a_{i}^{k}\left(\frac{r}{a}\right)^{i} , 
\end{equation}
which takes different coefficients for $u_{k}(r)$ and $v_{k}(r)$.
After getting eigenvalues and wave functions for the $0$-th order solutions, we
use those wave functions to calculate $p/m_Q$ corrections in order by order.
Corrections are estimated by evaluating matrix elements of corresponding terms
in the higher order Hamiltonian. Details of the prescription for carrying out
numerical calculations and getting reliable solutions are described in
\cite{Matsuki97}.

In the work of \cite{Matsuki97}, by fitting the smallest eigenvalues of the
Hamiltonian with masses of $D(1867)$ and $D^*(2008)$ for $c\bar q$ ($q=u, d$),
$D_s(1969)$ and $D_s^*(2112)$ for $c\bar s$, and $B(5279)$ and $B(5325)$ for
$b\bar q$, a strong coupling constant $\alpha_s$ and other potential parameters,
$a$ and $b$, were determined. At that time following the paper \cite{Morishita88},
we started the search for a set of parameters with $b<0$. Using those parameters
obtained this way, other mass levels were calculated and compared with the
experimental data for $D_{(s)}/B_{(s)}$ mesons. Light quark ($u$ and $d$) mass
was taken to be 10MeV as an input, being close to a value of current quark mass.

Here in this paper since we have now a plenty of the observed data, we follow a
different way of analysis.
\def\labelenumi{\theenumi)}
\begin{enumerate}
\item~First, we determine the six parameters, i.e., a
strong coupling constant $\alpha_s^c$, potential parameters $a$ and $b$, quark
masses $m_{u,d}$\footnote{In the numerical analysis, we set $m_{u}=m_{d}$ in
order to reduce the number of parameters.}, $m_s$, and $m_c$ by fitting
calculations to the observed six $D$ meson masses, i.e., those of two $\ell=0$
$(0^-, 1^-)$ states and four $\ell=1$ $(0^+, 1^+, 1^+, 2^+)$ states, and
similarly six $D_s$ meson masses, using the Minuit $\chi^2$ analysis.
The results are given in Table \ref{parameter}.
As written in the Introduction, we have found that the optimal value for $b$
becomes positive contrary to the former paper \cite{Matsuki97}, which affects
all other parameter values as well as the calculated values, $M_{\rm calc}$,
$M_0$, $p_i$, $n_i$, and $c_i$ whose meanings are explained below.
\item~Then, using the optimal parameters obtained this way except for a strong
coupling,
the bottom quark mass $m_b$ and a strong coupling $\alpha_s^b$ which is assumed
to be different from $\alpha_s^c$ for $D/D_s$ mesons are
determined by fitting the four observed $B$ meson masses and two $B_s$ meson masses.
The most optimal values of parameters are presented in Table \ref{parameter} at the
first order calculation in $p/m_Q$.
\end{enumerate}

\begin{table*}[t!]
\caption{Most optimal values of parameters.}
\label{parameter}
\begin{tabular}{lcccccccc}
\hline
\hline
Parameters
& ~~$\alpha_s^c$ & ~~$\alpha_s^b$ & ~~$a$ (GeV$^{-1}$) & ~~$b$ (GeV) 
& ~~$m_{u, d}$ (GeV) & ~~$m_s$ (GeV) & ~~$m_c$ (GeV) & ~~$m_b$ (GeV) \\
& ~~0.261 & ~~0.393 & ~~1.939 & ~~0.0749 
& ~~0.0112 & ~~0.0929 & ~~1.032 & ~~4.639 \\
\hline
\hline
\end{tabular}
\end{table*}

The masses with the same value of $k$ degenerate in the $0$-th order for
members of each spin doublet, are labeled as $M_0$. $c_i$ denotes the $i$-th
order correction in $p/m_Q$ expansion (in this paper $n=1, 2$) and thus, the
calculated heavy meson mass $M_{\rm calc}$ is given by the sum of $M_0$ and
corrections up to the $n$-th order,
\begin{equation}
M_{\rm calc} = M_0 + \sum_{i=1}^{n}\left(p_i+n_i\right) ,
\end{equation}
where $p_i$ includes the contributions stemming from only upper
components of a heavy quark, $n_i$ from lower, and $c_i=p_i+n_i$ in each
order. Here one should notice that $M_0=m_Q+E_0$.
The degeneracy in $k$ or for the states in the same spin doublet is resolved by
taking account of higher order corrections in $p/m_Q$ expansion. See Fig. \ref{mass_level}.

In our formalism, each state is uniquely classified by two quantum numbers,
$k$ and $j$, and
also approximately classified by the upper component of the wave function of
Eq.(\ref{0thsols2}) in terms of the conventional notation $^{2s+1}\ell_j$.
In terms of this notation, the $j^P =1^+$ state is a mixed state of $^3P_1$ and
$^1P_1$, while the $j^P =2^-$ state is a mixed state of $^3D_2$ and $^1D_2$. We
approximately denote them with double quotations in Tables
\ref{Dmeson}--\ref{Bsmeson}. The relations between them are given by
\cite{Matsuki97}
\begin{equation}
\left(
\begin{array}{c}
|"^3P_1"\rangle \\ |"^1P_1"\rangle
\end{array}
\right)
=\frac{1}{\sqrt{3}}
\left(
\begin{array}{cc}
\sqrt{2} & 1 \\ -1 & \sqrt{2}
\end{array}
\right)
\left(
\begin{array}{c}
|^3P_1\rangle \\ |^1P_1\rangle
\end{array}
\right) ,
\end{equation}
\begin{equation}
\left(
\begin{array}{c}
|"^3D_2"\rangle \\ |"^1D_2"\rangle
\end{array}
\right)
=\frac{1}{\sqrt{5}}
\left(
\begin{array}{cc}
\sqrt{3} & \sqrt{2} \\ -\sqrt{2} & \sqrt{3}
\end{array}
\right)
\left(
\begin{array}{c}
|^3D_2\rangle \\ |^1D_2\rangle
\end{array}
\right) .
\end{equation}

\subsection{$D$ and $D_s$}
\begin{figure}[t]
\includegraphics[scale=0.5,clip]{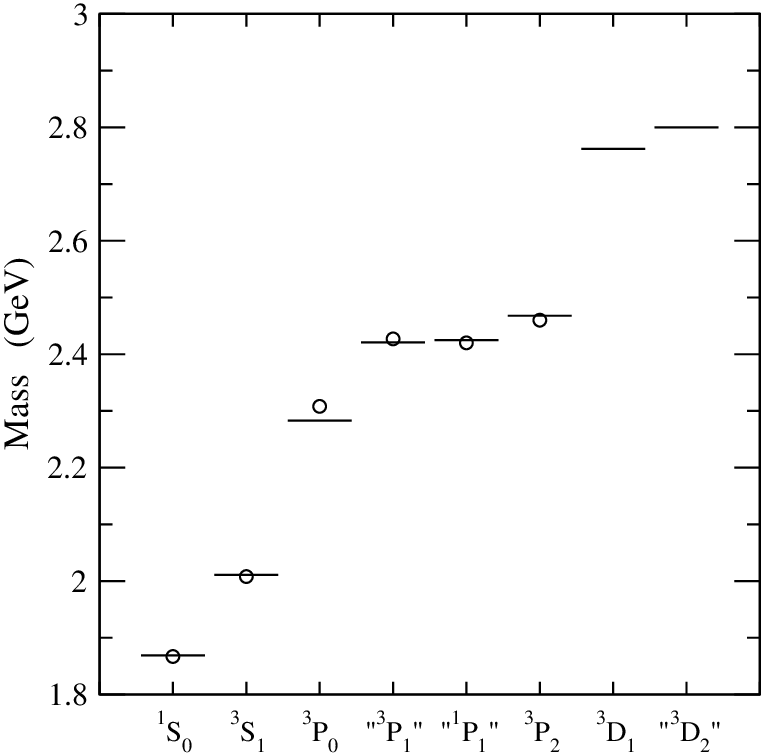}
\caption{Plot of $D$ meson masses in GeV. 
Solid bars and white circles represent calculated and observed masses, respectively. 
Specific values are given in Table \ref{Dmeson}. }
\label{fig-D}
\end{figure}

\begin{figure}[t]
\includegraphics[scale=0.5,clip]{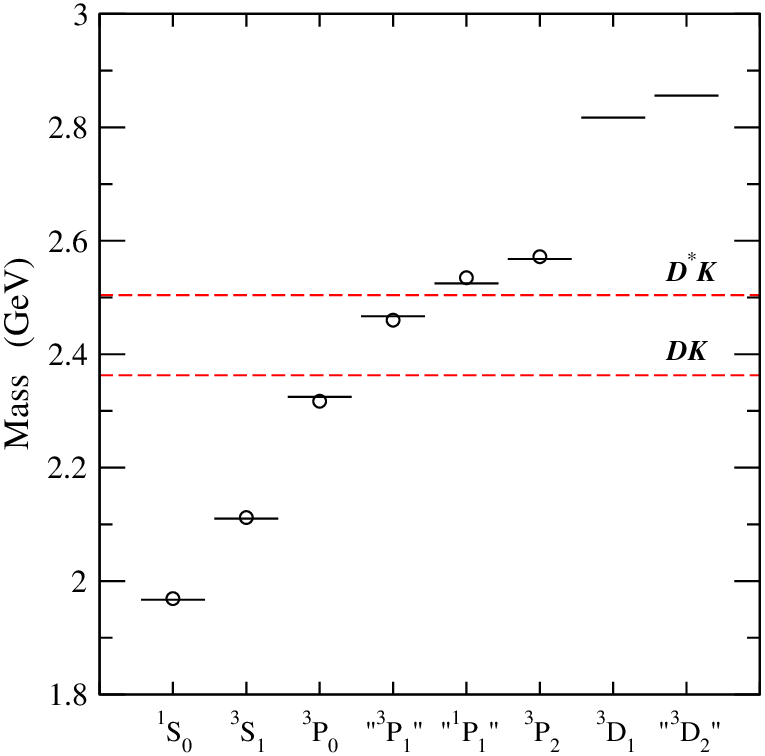}
\caption{Plot of $D_s$ meson masses. 
The legend is as for Fig. \ref{fig-D}. 
The dashed lines show $DK$ and $D^{*}K$ thresholds, respectively. 
Specific values are given in Table \ref{Dsmeson}.}
\label{fig-Ds}
\end{figure}

Calculated and optimally fitted values $M_{\rm calc}$ of $D$ and $D_s$ meson
masses with the first order corrections are presented in Tables \ref{Dmeson}
and \ref{Dsmeson}, respectively, together with experimental data $M_{\rm obs}$.
Those mass spectra of $D$ and $D_s$ mesons are also plotted in Figs.
\ref{fig-D} and \ref{fig-Ds} and Figure \ref{fig-Ds} also includes
$DK/D^{*}K$ threshold lines.
As one can see easily, our calculated masses $M_{\rm calc}$ are in good
agreement with each observed value $M_{\rm obs}$, and the discrepancies are
less than $1\%$. Especially it is remarkable that newly observed levels
$^3P_0(0^+)$ and $"^3P_1(1^+)"$ of both $D$ and $D_s$ mesons are reproduced
well, where the masses of $D_{sJ}$ mesons are below $DK/D^{*}K$ thresholds.
These levels of $D$ and $D_s$ mesons cannot be reproduced by any other quark
potential models
which predict about $100\sim 200$ MeV higher than our values even though they
also succeed in reproducing other levels, i.e., $^1S_0(0^-)$, $^3S_1(1^-)$,
$"^1P_1(1^+)"$, and $^3P_2(2^+)$.\footnote{See Table XIV of \cite{Swanson06}.}
This result gives us a great confidence that our framework may give a good solution to the narrow $D_{sJ}$ puzzle.

\begin{table*}[t!]
\caption{$D$ meson mass spectra (units are in MeV).}
\label{Dmeson}
\begin{tabular}{@{\hspace{0.5cm}}c@{\hspace{0.5cm}}c@{\hspace{1cm}}r@{\hspace{1cm}}r@{\hspace{1cm}}r@{\hspace{1cm}}c@{\hspace{1cm}}c@{\hspace{0.5cm}}}
\hline
\hline
$^{2s+1}L_J (J^P)$ & $M_0$ & 
\multicolumn{1}{c@{\hspace{1cm}}}{$c_1 /M_0$} & 
\multicolumn{1}{c@{\hspace{1cm}}}{$p_1 /M_0$} & 
\multicolumn{1}{c@{\hspace{1cm}}}{$n_1 /M_0$} & 
$M_{\rm calc}$ & $M_{\rm obs}$ \\
\hline
\multicolumn{1}{@{\hspace{0.6cm}}l}{$^1S_0 (0^-)$} 
& 1784 & 0.476 $\times 10^{-1}$ 
& 0.374 $\times 10^{-1}$ & 1.013 $\times 10^{-2}$ 
& 1869 & 1867 \\
\multicolumn{1}{@{\hspace{0.6cm}}l}{$^3S_1 (1^-)$} 
&  & 1.271 $\times 10^{-1}$ 
& 1.266 $\times 10^{-1}$ & 0.512 $\times 10^{-3}$ 
& 2011 & 2008 \\
\multicolumn{1}{@{\hspace{0.6cm}}l}{$^3P_0 (0^+)$} 
& 2067 & 1.046 $\times 10^{-1}$ 
& 0.959 $\times 10^{-1}$ & 0.874 $\times 10^{-2}$ 
& 2283 & 2308 \\
\multicolumn{1}{@{\hspace{0.6cm}}l}{$"^3P_1" (1^+)$} 
&  & 1.713 $\times 10^{-1}$ 
& 1.689 $\times 10^{-1}$ & 2.444 $\times 10^{-3}$ 
& 2421 & 2427 \\
\multicolumn{1}{@{\hspace{0.6cm}}l}{$"^1P_1" (1^+)$} 
& 2125 & 1.415 $\times 10^{-1}$ 
& 1.410 $\times 10^{-1}$ & 0.486 $\times 10^{-3}$ 
& 2425 & 2420 \\
\multicolumn{1}{@{\hspace{0.6cm}}l}{$^3P_2 (2^+)$} 
&  & 1.618 $\times 10^{-1}$ 
& 1.617 $\times 10^{-1}$ & 1.364 $\times 10^{-4}$ 
& 2468 & 2460 \\
\multicolumn{1}{@{\hspace{0.6cm}}l}{$^3D_1 (1^-)$} 
& 2322 & 1.894 $\times 10^{-1}$ 
& 1.872 $\times 10^{-1}$ & 2.228 $\times 10^{-3}$ 
& 2762 & $-$ \\
\multicolumn{1}{@{\hspace{0.6cm}}l}{$"^3D_2" (2^-)$} 
&  & 2.054 $\times 10^{-1}$ 
& 2.052 $\times 10^{-1}$ & 1.248 $\times 10^{-4}$ 
& 2800 & $-$ \\
\hline
\hline
\end{tabular}
\end{table*}
\begin{table*}[t!]
\caption{$D_s$ meson mass spectra (units are in MeV).}
\label{Dsmeson}
\begin{tabular}{@{\hspace{0.5cm}}c@{\hspace{0.5cm}}c@{\hspace{1cm}}
r@{\hspace{1cm}}r@{\hspace{1cm}}r@{\hspace{1cm}}c@{\hspace{1cm}}c@{\hspace{0.5cm}}}
\hline
\hline
$^{2s+1}L_J (J^P)$ & $M_0$ & 
\multicolumn{1}{c@{\hspace{1cm}}}{$c_1 /M_0$} & 
\multicolumn{1}{c@{\hspace{1cm}}}{$p_1 /M_0$} & 
\multicolumn{1}{c@{\hspace{1cm}}}{$n_1 /M_0$} & 
$M_{\rm calc}$ & $M_{\rm obs}$ \\
\hline
\multicolumn{1}{@{\hspace{0.6cm}}l}{$^1S_0 (0^-)$} 
& 1900 & 0.352 $\times 10^{-1}$ 
& 0.270 $\times 10^{-1}$ & 0.816 $\times 10^{-2}$ 
& 1967 & 1969 \\
\multicolumn{1}{@{\hspace{0.6cm}}l}{$^3S_1 (1^-)$} 
&  & 1.102 $\times 10^{-1}$ 
& 1.098 $\times 10^{-1}$ & 4.076 $\times 10^{-4}$ 
& 2110 & 2112 \\
\multicolumn{1}{@{\hspace{0.6cm}}l}{$^3P_0 (0^+)$} 
& 2095 & 1.101 $\times 10^{-1}$ 
& 1.027 $\times 10^{-1}$ & 0.740 $\times 10^{-2}$ 
& 2325 & 2317 \\
\multicolumn{1}{@{\hspace{0.6cm}}l}{$"^3P_1" (1^+)$} 
&  & 1.779 $\times 10^{-1}$ 
& 1.752 $\times 10^{-1}$ & 2.620 $\times 10^{-3}$ 
& 2467 & 2460 \\
\multicolumn{1}{@{\hspace{0.6cm}}l}{$"^1P_1" (1^+)$} 
& 2239 & 1.274 $\times 10^{-1}$ 
& 1.270 $\times 10^{-1}$ & 3.860 $\times 10^{-4}$ 
& 2525 & 2535 \\
\multicolumn{1}{@{\hspace{0.6cm}}l}{$^3P_2 (2^+)$} 
&  & 1.467 $\times 10^{-1}$ 
& 1.466 $\times 10^{-1}$ & 1.035 $\times 10^{-4}$ 
& 2568 & 2572 \\
\multicolumn{1}{@{\hspace{0.6cm}}l}{$^3D_1 (1^-)$} 
& 2342 & 2.032 $\times 10^{-1}$ 
& 2.008 $\times 10^{-1}$ & 2.382 $\times 10^{-3}$ 
& 2817 & $-$ \\
\multicolumn{1}{@{\hspace{0.6cm}}l}{$"^3D_2" (2^-)$} 
&  & 2.196 $\times 10^{-1}$ 
& 2.195 $\times 10^{-1}$ & 0.989 $\times 10^{-4}$ 
& 2856 & $-$ \\
\hline
\hline
\end{tabular}
\end{table*}

The first order corrections for $D$ and $D_s$ mesons are moderately large,
whose amount is an order of $10\%$. Since the charm quark mass in our fitting
is rather small, second order corrections might be necessary to get more
reliable results. However, we have found that they are only an order of $1\%$
or less and thus have neglected them in this work.
To demonstrate this smallness, we also present the fit for $D$ meson with the
second order corrections in Table \ref{Dmeson2nd} as a typical example. When
one looks at this Table, one notices that the fit with the experiments does not
increase accuracy compared with Table \ref{Dmeson} of the first order.
Hence from
here on in our computation we use only the first order calculations to obtain
any levels including radial excitations which we present in a separate paper
\cite{Matsuki06}.

\begin{table*}[b]
\caption{Second order $D$ meson mass spectra (units are in MeV).}
\label{Dmeson2nd}
\begin{tabular}{@{\hspace{0.5cm}}c@{\hspace{1cm}}c@{\hspace{1cm}}
r@{\hspace{1cm}}r@{\hspace{1cm}}c@{\hspace{1cm}}c@{\hspace{0.5cm}}}
\hline
\hline
$^{2s+1}\ell_j~(j^P)$ & $M_0$ & 
\multicolumn{1}{@{\hspace{0.6cm}}l}{$c_1 /M_0$} & 
\multicolumn{1}{@{\hspace{0.6cm}}l}{$c_2 /M_0$} & 
$M_{\rm calc}$ & $M_{\rm obs}$ \\
\hline
\multicolumn{1}{@{\hspace{0.6cm}}l}{$^1S_0(0^-)$} 
& 1783 & 0.678 $\times 10^{-1}$ &-2.091 $\times 10^{-2}$ & 1867 & 1867 \\
\multicolumn{1}{@{\hspace{0.6cm}}l}{$^3S_1(1^-)$} 
& & 1.245 $\times 10^{-1}$ & 2.507 $\times 10^{-3}$ & 2009 & 2008 \\
\multicolumn{1}{@{\hspace{0.6cm}}l}{$^3P_0(0^+)$} 
& 1935 & 1.694 $\times 10^{-1}$ & 1.567 $\times 10^{-2}$ & 2293 & 2308 \\
\multicolumn{1}{@{\hspace{0.6cm}}l}{$"^3P_1"(1^+)$} 
& & 2.132 $\times 10^{-1}$ & 1.665 $\times 10^{-3}$ & 2350 & 2427 \\
\multicolumn{1}{@{\hspace{0.6cm}}l}{$"^1P_1"(1^+)$} 
& 2045 & 1.412 $\times 10^{-1}$ & 2.499 $\times 10^{-3}$ & 2432 & 2420 \\
\multicolumn{1}{@{\hspace{0.6cm}}l}{$^3P_2(2^+)$} 
& & 1.616 $\times 10^{-1}$ & -1.003 $\times 10^{-3}$ & 2448 & 2460 \\
\multicolumn{1}{@{\hspace{0.6cm}}l}{$^3D_1(1^-)$} 
& 2127 & 1.863 $\times 10^{-1}$ & 2.109 $\times 10^{-4}$ & 2803 & $-$ \\
\multicolumn{1}{@{\hspace{0.6cm}}l}{$"^3D_2"(2^-)$} 
& & 2.021 $\times 10^{-1}$ & -1.409 $\times 10^{-3}$ & 2726 & $-$ \\
\hline
\hline
\end{tabular}
\end{table*}

\subsection{$B$ and $B_s$}
\begin{figure}[t]
\includegraphics[scale=0.5,clip]{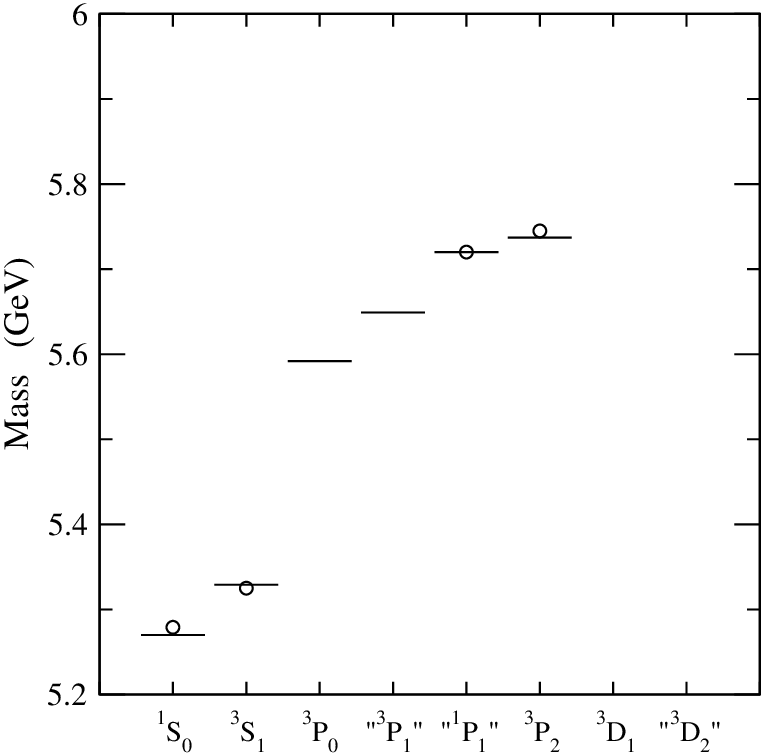}
\caption{Plot of $B$ meson masses. 
The legend is as for Fig. \ref{fig-D}. 
Specific values are given in Table \ref{Bmeson}.}
Calculated mass values for $^3D_1$ and "$^3D_2$" are too large to
put them in the same figure with others.
\label{fig-B}
\end{figure}

\begin{figure}[t]
\includegraphics[scale=0.5,clip]{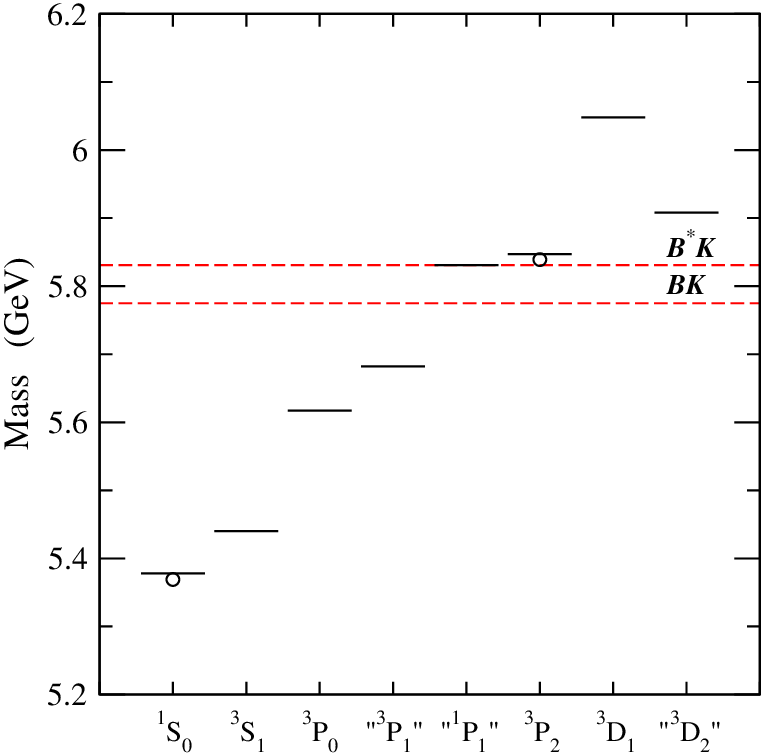}
\caption{Plot of $B_s$ meson masses. 
The legend is as for Fig. \ref{fig-D}. 
The dashed lines show $BK$ and $B^{*}K$ thresholds, respectively. 
Specific values are given in Table \ref{Bsmeson}.}
\label{fig-Bs}
\end{figure}

For $B$ and $B_s$ mesons, unfortunately there are only a few data available
although newly discovered states ($B_1(5720)$, $B_2^*(5745)$, and
$B_{s2}^*(5839)$) are reported recently by D0 and CDF.\cite{CDF_D006}
Calculated masses of $B$ and $B_s$ mesons are given in Tables \ref{Bmeson} and
\ref{Bsmeson}, respectively. We have used a new value of a strong coupling
$\alpha_s^b$ given by Table \ref{parameter} for calculating $B$ and $B_s$ mesons
different from $\alpha_s^c$ for $D$ meson in Table \ref{parameter}.
Newly discovered levels are well reproduced by our model, too, as shown in these
Tables. We predict the mass of several excited states which are not yet observed.
The mass spectra of $B$ and $B_s$ mesons are also plotted in Figs. \ref{fig-B}
and \ref{fig-Bs} and Figure \ref{fig-Bs} also includes $BK/B^{*}K$
threshold lines. Among these levels, predicted masses of $0^+$ and $1^+$ states
for $B_s$ mesons are below $BK/B^{*}K$ threshold the same as the case for
$D_{sJ}$ mesons. Therefore, their decay modes are kinematically forbidden, and
the dominant decay modes are the pionic decay which is the same results obtained in
\cite{Matsuki05} although values of $B_s$ mesons are slightly different from each
other:
\begin{subequations}
\begin{eqnarray}
B_s (0^{+}) &\to& B_s (0^{-})+\pi , \\
B_s (1^{+}) &\to& B_s (1^{-})+\pi .
\end{eqnarray}
\label{Bs-decay}
\end{subequations}
The decay widths of these states are expected to be narrow the same as $D_{sJ}$
meson cases, since those decay modes of Eq.(\ref{Bs-decay}) violate the isospin
invariance. These higher states might be observed in Tevatron/LHC experiments or
even in D0 and/or CDF of FNAL in the near future by analyzing the above decay modes.
This is because D0 and CDF have recently announced the discovery of
$B_1(5720)=B("^1P_1(1^+)")$, $B_2^*(5745)=B(^3P_2(2^+))$, and
$B_{s2}^*(5839)=B_s(^3P_2(2^+))$, which have narrow decay width with the same decay
products as in Eq.(\ref{Bs-decay}) since only the D-wave decay channel is
possible.\cite{CDF_D006} Even though these mass values are reproduced by other
models \cite{Eichten93, Pierro01, Ebert98, Orsland99}, but again those models
predict larger mass values for $0^+$ and $1^+$ states of $B_s$ mesons
than ours, the same situation as in $D_{sJ}$ so that these states have broad decay
width contrary to our prediction. We hope that our prediction could be
tested in the forthcoming experiments which distinguish ours from other models.

\begin{table*}[t!]
\caption{$B$ meson mass spectra (units are in MeV).}
\label{Bmeson}
\begin{tabular}{@{\hspace{0.5cm}}c@{\hspace{0.5cm}}c@{\hspace{1cm}}
r@{\hspace{1cm}}r@{\hspace{1cm}}r@{\hspace{1cm}}c@{\hspace{1cm}}c@{\hspace{0.5cm}}}
\hline
\hline
$^{2s+1}L_J (J^P)$ & $M_0$ & 
\multicolumn{1}{c@{\hspace{1cm}}}{$c_1 /M_0$} & 
\multicolumn{1}{c@{\hspace{1cm}}}{$p_1 /M_0$} & 
\multicolumn{1}{c@{\hspace{1cm}}}{$n_1 /M_0$} & 
$M_{\rm calc}$ & $M_{\rm obs}$ \\
\hline
\multicolumn{1}{@{\hspace{0.6cm}}l}{$^1S_0 (0^-)$} 
& 5277 & -0.161 $\times 10^{-2}$ 
& -3.795 $\times 10^{-3}$ & 2.187 $\times 10^{-3}$ 
& 5270 & 5279 \\
\multicolumn{1}{@{\hspace{0.6cm}}l}{$^3S_1 (1^-)$} 
&  & 0.981 $\times 10^{-2}$ 
& 9.706 $\times 10^{-3}$ & 1.107 $\times 10^{-4}$ 
& 5329 & 5325 \\
\multicolumn{1}{@{\hspace{0.6cm}}l}{$^3P_0 (0^+)$} 
& 5570 & 0.401 $\times 10^{-2}$ 
& 1.937 $\times 10^{-3}$ & 2.072 $\times 10^{-3}$ 
& 5592 & $-$ \\
\multicolumn{1}{@{\hspace{0.6cm}}l}{$"^3P_1" (1^+)$} 
&  & 1.412 $\times 10^{-2}$ 
& 1.400 $\times 10^{-2}$ & 1.227 $\times 10^{-4}$ 
& 5649 & $-$ \\
\multicolumn{1}{@{\hspace{0.6cm}}l}{$"^1P_1" (1^+)$} 
& 5660 & 1.069 $\times 10^{-2}$ 
& 1.066 $\times 10^{-2}$ & 0.289 $\times 10^{-4}$ 
& 5720 & 5720 \\
\multicolumn{1}{@{\hspace{0.6cm}}l}{$^3P_2 (2^+)$} 
&  & 1.364 $\times 10^{-2}$ 
& 1.364 $\times 10^{-2}$ & 2.120 $\times 10^{-7}$ 
& 5737 & 5745 \\
\multicolumn{1}{@{\hspace{0.6cm}}l}{$^3D_1 (1^-)$} 
& 5736 & 2.203 $\times 10^{-1}$ 
& 2.202 $\times 10^{-1}$ & 4.583 $\times 10^{-5}$ 
& 6999 & $-$ \\
\multicolumn{1}{@{\hspace{0.6cm}}l}{$"^3D_2" (2^-)$} 
&  & 1.430 $\times 10^{-1}$ 
& 1.430 $\times 10^{-1}$ & 2.092 $\times 10^{-7}$ 
& 6556 & $-$ \\
\hline
\hline
\end{tabular}
\end{table*}
\begin{table*}[t!]
\caption{$B_s$ meson mass spectra (units are in MeV).}
\label{Bsmeson}
\begin{tabular}{@{\hspace{0.5cm}}c@{\hspace{0.5cm}}c@{\hspace{1cm}}
r@{\hspace{1cm}}r@{\hspace{1cm}}r@{\hspace{1cm}}c@{\hspace{1cm}}c@{\hspace{0.5cm}}}
\hline
\hline
$^{2s+1}L_J (J^P)$ & $M_0$ & 
\multicolumn{1}{c@{\hspace{1cm}}}{$c_1 /M_0$} & 
\multicolumn{1}{c@{\hspace{1cm}}}{$p_1 /M_0$} & 
\multicolumn{1}{c@{\hspace{1cm}}}{$n_1 /M_0$} & 
$M_{\rm calc}$ & $M_{\rm obs}$ \\
\hline
\multicolumn{1}{@{\hspace{0.6cm}}l}{$^1S_0 (0^-)$} 
& 5394 & -0.302 $\times 10^{-2}$ 
& -0.485 $\times 10^{-2}$ & 0.183 $\times 10^{-2}$ 
& 5378 & 5369 \\
\multicolumn{1}{@{\hspace{0.6cm}}l}{$^3S_1 (1^-)$} 
&  & 0.853 $\times 10^{-2}$ 
& 0.844 $\times 10^{-2}$ & 9.249 $\times 10^{-5}$ 
& 5440 & $-$ \\
\multicolumn{1}{@{\hspace{0.6cm}}l}{$^3P_0 (0^+)$} 
& 5598 & 0.350 $\times 10^{-2}$ 
& 0.173 $\times 10^{-2}$ & 0.177 $\times 10^{-2}$ 
& 5617 & $-$ \\
\multicolumn{1}{@{\hspace{0.6cm}}l}{$"^3P_1" (1^+)$} 
&  & 1.498 $\times 10^{-2}$ 
& 1.444 $\times 10^{-2}$ & 5.396 $\times 10^{-4}$ 
& 5682 & $-$ \\
\multicolumn{1}{@{\hspace{0.6cm}}l}{$"^1P_1" (1^+)$} 
& 5775 & 0.978 $\times 10^{-2}$ 
& 0.969 $\times 10^{-2}$ & 8.257 $\times 10^{-5}$ 
& 5831 & $-$ \\
\multicolumn{1}{@{\hspace{0.6cm}}l}{$^3P_2 (2^+)$} 
&  & 1.263 $\times 10^{-2}$ 
& 1.261 $\times 10^{-2}$ & 2.014 $\times 10^{-5}$ 
& 5847 & 5839 \\
\multicolumn{1}{@{\hspace{0.6cm}}l}{$^3D_1 (1^-)$} 
& 5875 & 2.949 $\times 10^{-2}$ 
& 2.898 $\times 10^{-2}$ & 5.104 $\times 10^{-4}$ 
& 6048 & $-$ \\
\multicolumn{1}{@{\hspace{0.6cm}}l}{$"^3D_2" (2^-)$} 
&  & 0.564 $\times 10^{-2}$ 
& 0.562 $\times 10^{-2}$ & 1.980 $\times 10^{-5}$ 
& 5908 & $-$ \\
\hline
\hline
\end{tabular}
\end{table*}
%

\section{Conclusion and Discussion}
\label{summary}
Given recent new data and expanding a parameter space with positive $b$,
we have reanalyzed the mass spectra of heavy mesons, $D$, $D_s$, $B$ and $B_s$
with our semirelativistic quark potential model, in which an effective
Hamiltonian has both a heavy quark symmetry and a chiral symmetry in a certain
limit of parameters. Hamiltonian, wave function, and eigenvalue are all
consistently expanded in $p/m_Q$, and a light antiquark is treated as four-spinor
Dirac particle while a heavy quark is nonrelativistically reduced using the FWT
transformation.
Calculated masses of not only newly observed excited ($\ell=1$) states,
$D_{s0}(2317)$, $D_{s1}(2460)$, $D_{0}^{*}(2308)$ and $D_{1}'(2427)$ which are
not explained by conventional quark potential models, but also already
established ($\ell=0$ and $\ell=1$) states of $D$ and $D_s$ mesons are
simultaneously reproduced in good agreement with experimental data within one
percent of accuracy. Now we also have plausible
values of masses for $^3D_1$
and $"^3D_2"$ which we failed to give in the former paper \cite{Matsuki97}.
This has been achieved by finding the most optimal parameter set with positive
parameter $b$.

These results suggest that our model could be a good solution to the narrow
$D_{sJ}$ puzzle, which is considered to be still challenging and people are
still looking for exotic state possibilities. We would say that
the quark potential model still remains powerful enough to understand the heavy
meson spectroscopy. The important thing in a heavy-light system is to expand all
quantities in the model in $p/m_Q$ and set equations consistently order by order
and not to neglect the lower components of a heavy quark, which
has not been taken into account by other quark potential models. We have
also predicted the masses of higher excited states of $B$ and $B_s$ mesons
which are not yet observed, at the first order in $p/m_b$.  Among these, the
newly discovered states by D0 and CDF are also reproduced by our model, too.
There are other models to reproduce these states \cite{Eichten93, Pierro01,
Ebert98, Orsland99} but they cannot predict $0^+$ and $1^+$ states less than
$BK/B^{*}K$ threshold. This is the main difference of the results when
comparing our model with others. Other models for heavy mesons cannot yield the
masses of $^3P_0(0^+)$ and $"^3P_1(1^+)"$ states of $D_s$ and $B_s$ mesons much
lower than $"^1P_1(1^+)"$ and $^3P_2(2^+)$ states like ours.

Some comments are in order.

(i) As shown in Tables \ref{Dmeson}--\ref{Bsmeson}, the first order corrections
to the $0$-th order mass, $c_{1}/M_0$, are roughly an order of $10\%$ for $D$
and $D_s$ mesons, while $1\%$ for $B$ and $B_s$ mesons. Therefore, $p/m_Q$
expansion works much better for $B$ and $B_s$ mesons. This is quite natural,
since the bottom quark mass is heavier than the charm quark mass. In addition,
it is interesting that the first order correction becomes larger for higher
excited states. It means that the relativistic correction must be more
important in higher excited states, since the higher order corrections in
$p/m_Q$ expansion represent the relativistic effects of a heavy quark. This
is consistent with a naive picture that the inner motion of light as well as
heavy quarks may be larger in excited states, and therefore the effects of
lower components of a heavy quark may become much more important to be included.

(ii) The mass difference between members of a spin doublet satisfies the following
relation (given by Eq.(53) of \cite{Matsuki97}):
\begin{eqnarray}
  m_{c}(M_{D^{*}}-M_{D}) &=& m_{b}(M_{B^{*}}-M_{B}), \label{spin_doublet1} \\
  m_{c}(M_{D^{*}_s}-M_{D_s}) &=& m_{b}(M_{B^{*}_s}-M_{B_s}),
  \label{spin_doublet2}
\end{eqnarray}
in the first order of calculation. This is because each of two states having
the same value of the quantum number $k$ has an equal mass at the $0$-th order
and hence the mass is degenerate at this stage as shown in
Fig. \ref{mass_level}. The splitting occurs by including the first order
($p/m_Q$) corrections originated from $H_1$ in Eq.(\ref{H_mQ}), which is
proportional to $f^k(m_q)/m_Q$ with the same functional form $f^k(m_q)$ for two
states, and therefore Eqs.(\ref{spin_doublet1}, \ref{spin_doublet2}) are
exactly satisfied in our model. Similar relations hold for higher spin states
with the same value of $k$.

(iii) As shown in Table \ref{parameter}, the light quark masses $m_{u, d}$ and
$m_s$ are considerably small, $m_{u, d}\sim 10$ MeV and $m_{s}\sim 90$ MeV,
compared with the constituent quark masses which are adopted in
conventional potential models \cite{Godfrey85}.  It should be noted that these
values are not input but the outcome of this analysis, though in the previous
analysis \cite{Matsuki97}, $m_{u,d}=10$ MeV is taken as an input.
This result is consistent with those obtained in
\cite{Politzer, Matsuki80, Kaburagi}.

This situation might be described as follows. A light quark is moving around a
heavy quark in a heavy meson wearing small gluon clouds with large momenta in
our treatment of quarks so that light quarks have so-called current quark mass.
This holds only in heavy-light systems and in the other systems, e.g., quarkonium
and ordinary hadrons, we should use quarks wearing more gluon clouds, i.e.,
constituent quarks as adopted in usual quark potential models. There may be
another interpretation bridging our model with light mass with a constituent quark
model.

(iv) Repeatedly saying, the main difference between other conventional
quark potential models and ours is that our model systematically
expands interaction terms in $p/m_Q$, takes into account lower component
contributions of heavy quarks in the intermediate states when calculating
higher order terms, and adopt very light antiquark mass. With regard to $1/m_Q$
expansion, for instance, the model
of \cite{Godfrey85,Godfrey91} has kinetic terms, $\sqrt{p^2+m^2}$, while they
adopted nonrelativistic interaction terms which are symmetric in $Q$ and
$\bar q$ or in $m_Q$ and $m_{\bar q}$.
The model of \cite{Ebert98} somehow manipulated to hold heavy quark symmetry
and expanded interactions in $1/m_Q$ but adopted constituent masses for light
quarks, $m_u$, $m_d$, and $m_s$. Even so, their numerical results give about a
hundred MeV larger values for $D_{sJ}$ particles.
Both models have not taken into account the lower components of a heavy
quark which naturally appear in our semirelativistic potential model.

(v) We have been successful in reproducing experimental data for both heavy
mesons $D/D_s$ and $B/B_s$. However we have also left unresolved problems in
our semirelativistic model. When one looks at Table \ref{parameter}, one notices
that values of a strong coupling used for $D/D_s$ and $B/B_s$ are largely
different from each other. Especially that for $D/D_s$ is too small to be
a strong coupling. Also we have not used a running strong coupling although
most of other quark potential models take into account this effect in some way.

\begin{acknowledgments}
One of the authors (K.S.) would like to thank S. Yokkaichi for 
helpful discussions about numerical calculations.
\end{acknowledgments}

\begin{appendix}
\section{Schr\"odinger equation with an Effective Hamiltonian}
\label{appen_Hamiltonian}
The original Schr\"odinger equation is actually modified by the FWT
transformation in our framework in order to make non-relativistic reduction of a
heavy quark. Now we recall the modified Schr\"odinger
equation of Eq. (\ref{eigen_FWT}), 
\begin{equation}
(H_{\rm FWT}-m_{Q}) \otimes\psi_{\rm FWT} =\tilde{E}\psi_{\rm FWT},
\end{equation}
The Hamiltonian and the wave function are transformed by the
following FWT and charge conjugate operator on the heavy quark sector:
\begin{eqnarray}
  H_{FWT}&=&U_{c}U_{FWT}(p'_{Q})HU^{-1}_{FWT}(p_{Q})U^{-1}_{c} , \\
  \psi_{FWT}&=&U_{c}U_{FWT}(p_{Q})\psi , 
\end{eqnarray}
where given are
\begin{eqnarray}
  U_{FWT}(p)&=&\exp\left(W(p)\vmb{\gamma}_{Q}\cdot
  \vec{\hat{\mbox{\boldmath$p$}}}\right)
  =\cos W(p) + \vmb{\gamma}_{Q}\cdot \vec{\hat{\mbox{\boldmath$p$}}}
  \sin W(p) , \\
  U_{c}&=& i\gamma^0_{Q}\gamma^2_{Q}=-U^{-1}_{c} , 
\end{eqnarray}
and some kinematical variables are defined as 
\begin{eqnarray}
  && \vec{\hat{\mbox{\boldmath$p$}}}=\frac{\vmb{p}}{p}, ~~~
  \tan W(p)=\frac{p}{m_{Q}+E}, ~~~E=\sqrt{\vmb{p}^{2}+m^2_{Q}} , \\
  && \vmb{p}=\vmb{p}_{q}=-\vmb{p}_{Q}, ~~~\vmb{p}'=\vmb{p}'_{q}=-\vmb{p}'_{Q} ,
  ~~~\vmb{q}=\vmb{p}'-\vmb{p}.
\end{eqnarray}
Here we are considering a scattering of a heavy quark $Q$ and a light antiquark
$\bar{q}$. Hence, $\vmb{p}_i$, $\vmb{p}_i'$, and $\vmb{q}$ given above
represent momenta of incoming, outgoing quarks, and gluon which is exchanged
between quarks, respectively. Note that the argument of the FWT transformation
$U_{FWT}$ operating on a Hamiltonian from left is different from the
right-operating one, since an outgoing momentum $\vmb{p}'_{Q}$ is different
from an incoming one $\vmb{p}_{Q}$. However, in our study we work in a
configuration space in which momenta are nothing but the derivative operators.
When we write them differently, for instance as $\vmb{p}_{Q}$ and
$\vmb{p}'_{Q}$, their expressions are reminders of their momentum
representation. 
Therefore, although the arguments of $U_{FWT}$ and $U^{-1}_{FWT}$ look
different, $\vmb{p}_{Q}$ and $\vmb{p}'_{Q}$ are expressed by the same derivative
operator $-i\vmb{\nabla}$. The momentum transfer $\vmb{q}$ operates only on
potentials and provides nonvanishing results. 

The charge conjugation operator $U_c$ is introduced to make the wave function
$U_{c}\psi$ a true bi-spinor, i.e., gamma matrices of a light antiquark are
multiplied from left while those of a heavy quark from right, which is
expressed by using a notation $\otimes$. 

Then, the Hamiltonian in the modified Schr\"odinger equation given by Eq.
(\ref{eigen_FWT}) is expanded in powers of $p/m_Q$: 
\begin{equation}
H_{\rm FWT}-m_{Q}=H_{-1}+H_{0}+H_{1}+H_{2}+\cdots . 
\end{equation}
$H_{i}$ stands for the $i$-th order expanded Hamiltonian whose explicit forms
are given by 
\begin{widetext}
\begin{subequations}
\begin{eqnarray}
H_{-1}&=&-(1+\beta_{Q})m_{Q} , 
\label{Hm1} \\
H_{0~}&=&\vmb{\alpha}_{q}\cdot\vmb{p} 
+ \beta_{q}m_{q} 
- \beta_{q}\beta_{Q}S 
+ \left\{1 + \frac{1}{2}\left[\vmb{\alpha}_{q}\cdot \vmb{\alpha}_{Q} +
(\vmb{\alpha}_{q}\cdot \vmb{n})(\vmb{\alpha}_{Q}\cdot \vmb{n}) \right]\right\} V , \\
H_{1~}&=&-\frac{1}{2m_{Q}}\beta_{Q}\vmb{p}^{2} 
+ \frac{1}{m_{Q}}\beta_{q}\vmb{\alpha}_{Q}\cdot
\left(\vmb{p}+\frac{1}{2}\vmb{q}\right)S 
+ \frac{1}{2m_{Q}}\vmb{\gamma}_{Q}\cdot \vmb{q}V \nonumber \\
&&- \frac{1}{2m_{Q}}\left[\beta_{Q}\left(\vmb{p}+\frac{1}{2}\vmb{q}\right) +
i\vmb{q}\times\beta_{Q}\vmb{\Sigma}_{Q}\right]
\cdot \left[\vmb{\alpha}_{q} + (\vmb{\alpha}_{q}\cdot \vmb{n})\vmb{n}\right] V , \\
H_{2~}&=&\frac{1}{2m_Q^2}\beta_{q}\beta_{Q}
\left(\vmb{p}+\frac{1}{2}\vmb{q}\right)^{2}S 
- \frac{i}{4m_Q^2}\vmb{q}\times \vmb{p}\cdot \beta_{q}\beta_{Q}\vmb{\Sigma}_{Q}S 
- \frac{1}{8m_Q^2}\vmb{q}^{2}V 
- \frac{i}{4m_Q^2}\vmb{q}\times\vmb{p}\cdot \vmb{\Sigma}_{Q}V \nonumber \\
&&- \frac{1}{8m_Q^2}\left\{(\vmb{p}+\vmb{q})(\vmb{\alpha}_{Q}\cdot\vmb{p})
+ \vmb{p}\left[\vmb{\alpha}_{Q}\cdot(\vmb{p}+\vmb{q})\right] 
+ i\vmb{q}\times\vmb{p}\gamma_Q^5 \right\}
\cdot \left[\vmb{\alpha}_{q}+(\vmb{\alpha}_{q}\cdot\vmb{n})\vmb{n}\right] V , \\
&& \vdots \nonumber 
\end{eqnarray}
\label{H_appen}
\end{subequations}
\end{widetext}
More details of matrix elements of each order in $1/m_Q$ and properties of
wave functions with several operators are given and evaluated in Ref.
\cite{Matsuki97}.

Eq.(\ref{Hm1}) means that the lowest order equation gives
$(1+\beta_Q)\psi=0$, which implies that a heavy quark is regarded as a
static source of color so that projection operator becomes
$(1\pm\beta_Q)/2$, the one at rest system. That is, the solution given by
Eq.(\ref{0thsols1}) includes only upper or positive components with regard
to a heavy quark. In this sense, upper/lower component of a heavy quark has the
same meaning as positive/negative energy component.


\font\singlebf=cmbx10 scaled\magstep1
\font\singlerm=cmr10 scaled\magstep1
\font\singless=cmss10 scaled\magstep1
\font\doublebf=cmbx10 scaled\magstep2
\font\doublerm=cmr10 scaled\magstep2
\font\triplebf=cmbx10 scaled\magstep3
\font\triplerm=cmr10 scaled\magstep3
\font\hsym=cmsy10
\font\singlehsym=cmsy10 scaled\magstep1
\font\sevenit=cmti7
\def\hana#1{\hbox{\hsym #1}}
\def\bighana#1{\hbox{\singlehsym #1}}
\def\yohaku{\hskip1000pt minus 1fill}
\def\tabrule{\noalign{\hrule}}
\section{Bloch Method}

This Appendix is based on the paper \cite{MS97}, by which
we explain Bloch's method \cite{Bloch} for a degenerate system. We will see the
Foldy-Wouthuysen-Tani \cite{Foldy50} transformation is automatically carried out
and will obtain the systematic expansion of the heavy-light system
in $1/m_Q$. To distinguish this formulation from the one we have used, we use
different notations in this Appendix.

     We expand the eigenvalues of the Hamiltonian with respect to the inverse of
$m_Q$.  First we divide the Hamiltonian as 
${\hana H}=\beta_Qm_Q+{\hana V}$, and we treat ${\hana V}$ as a perturbation,
where actually ${\hana H}=H$.
The eigenvalues of the unperturbed Hamiltonian are simply $\pm m_Q$, and the
eigenstates of ${\hana H}$ are almost degenerate.  Then we apply the
perturbation method of Bloch \cite{Bloch}.  In general, let us consider the
Hamiltonian 
${\hana H}=H_0+{\hana V}$.  We denote the projection operator to the subspace
$M$ with an eigenvalue $E$ of $H_0$ as $P$, and the projection operator to the
subspace ${\hana M}$ with eigenvalues $\{{\hana E}_j\}$ of 
${\hana H}$ as ${\hana P}$, where all ${\hana E}_j$ converge to $E$ in the
vanishing limit of the interaction ${\hana V}$.  Further we choose appropriate
vectors $\{\psi_j\}$ in the subspace $M$ such that their projections on
${\hana M}$ become the eigenvectors of ${\hana H}$, that is,
${\hana H}{\hana P}\psi_j={\hana E}_j{\hana P}\psi_j$.
If we define an operator ${\hana U}$ by ${\hana P}P={\hana U}P{\hana P}P$, then
we get the following eigenvalue equation,
\begin{equation}
 {\hana H}_{eff}(P{\hana P}\psi_j)=
{\hana E}_j(P{\hana P}\psi_j),
\end{equation}
where the effective Hamiltonian ${\hana H}_{eff}$ is defined by 
${\hana H}_{eff}=P{\hana H}{\hana U}$.  Here we must notice that this effective
Hamiltonian is not hermitian.  The eigenfunction of ${\hana H}$, ${\hana P}\psi_j$,
is obtained from the eigenfunction of 
${\hana H}_{eff}$ given by $\phi_j=P{\hana P}\psi_j$, by multiplying
${\hana U}$ from left, that is,
\begin{equation}
{\hana U}\phi_j={\hana U}P{\hana P}\psi_j={\hana P}\psi_j.
\end{equation}
Note that $P\psi_j=\psi_j$.
Again we must notice that ${\hana U}$ is not unitary, therefore, even if
$\phi_j$ is normalized, ${\hana U}\phi_j$ is not.

     ${\hana P}$ and ${\hana U}$ are formally expanded as follows.
\begin{equation}
{\hana P}=P-\sum_{m=1}^{\infty}\sum_{\{k_i\}}S_{k_1}{\hana V}S_{k_2}\cdots
S_{k_m}{\hana V}S_{k_{m+1}},
\end{equation}
where $S_k$'s are defined by 
\begin{equation}
S_0=-P\quad {\rm and} \quad S_k={Q\over a^k}\;\; (k=1,2,\cdots) 
\quad {\rm with} \quad Q=1-P,\quad a=E-H_0,
\end{equation}
and the sum over $\{k_i\}$ extends over non-negative integers satisfying the
condition 
$\sum_{i=1}^{m+1}k_i=m$.
\begin{equation}
{\hana U}=
\sum_{m=1}^{\infty}{\sum_{\{k_i\}}}'
S_{k_1}{\hana V}S_{k_2}\cdots S_{k_m}{\hana V}P,
\end{equation}
where the sum over $\{k_i\}$ extends over non-negative integers satisfying the
conditions 
$\sum_{i=1}^m k_i=m$ and
\par\noindent $\sum_{i=1}^p k_i\ge p\quad (p=1,2,\cdots m-1)$.  
Making use of the expansion formula for ${\hana U}$, we obtain the effective
Hamiltonian including the second order corrections in $1/a\sim 1/m_Q$ as follows:
\begin{equation}
{\hana H}_{eff}=EP+P{\hana V}P+P{\hana V}{Q\over a}{\hana V}P+
P{\hana V}{Q\over a}{\hana V}{Q\over a}{\hana V}P-
P{\hana V}{Q\over a^2}{\hana V}P{\hana V}P + \cdots.
\end{equation}
where $P=(1+\beta_Q)/2$ and $Q=(1-\beta_Q)/2$.

\end{appendix}

\def\Journal#1#2#3#4{{#1} {\bf #2}, #3 (#4)}
\def\NIM{Nucl. Instrum. Methods}
\def\NIMA{Nucl. Instrum. Methods A}
\def\NPB{Nucl. Phys. B}
\def\PLB{Phys. Lett. B}
\def\PRL{Phys. Rev. Lett.}
\def\PRD{Phys. Rev. D}
\def\PRO{Phys. Rev.}
\def\ZPC{Z. Phys. C}
\def\EPJ{Eur. Phys. J. C}
\def\PR{Phys. Rept.}
\def\IJM{Int. J. Mod. Phys. A}
\def\PTP{Prog. Theor. Phys.}

\end{document}